# Direct observation of the leakage current in epitaxial diamond Schottky barrier devices by conductive-probe atomic force microscopy and Raman imaging

**J Alvarez[1], M Boutchich[1], J P Kleider[1], T Teraji[2] and Y Koide[2]**

[1] Laboratoire de Génie Electrique de Paris; CNRS UMR 8507; Supélec ; Sorbonne Universités-UPMC Univ. Paris 06 ; Université Paris XI ; 11 rue Joliot-Curie Plateau de Moulon, 91192, Gif-sur-Yvette, France
[2] Wide Bandgap Materials Group, Optical and Electronic Materials Unit, National Institute for Materials Science (NIMS), 1-1 Namiki, Tsukuba, Ibaraki 305-0044, Japan

E-mail: jose.alvarez@lgep.supelec.fr

**Abstract.** The origin of the high leakage current measured in several vertical-type diamond Schottky devices is conjointly investigated by conducting probe atomic force microscopy (CP-AFM) and confocal micro-Raman/Photoluminescence (PL) imaging analysis.
Local areas characterized by a strong decrease of the local resistance (5-6 orders of magnitude drop) with respect to their close surrounding have been identified in several different regions of the sample surface. The same local areas, also referenced as electrical hot-spots, reveal a slightly constrained diamond lattice and three dominant Raman bands in the low-wavenumber region (590, 914 and 1040 $cm^{-1}$). These latter bands are usually assigned to the vibrational modes involving boron impurities and its possible complexes that can electrically act as traps for charge carriers. Local current-voltage measurements performed at the hot-spots point out a trap-filled-limited (TFL) current as the main conduction mechanism favoring the leakage current in the Schottky devices.

## 1. Introduction
In the area of wide band-gap semiconductors, diamond remains an attractive material due fundamentally to its exceptional electronic, optical and physical properties [1, 2]. These properties justify largely the research efforts that have been done at the level of the crystal growth to improve its quality in order to suit particular applications. In the field of high power electronics, diamond attracts attention because, inter alia, its high thermal conductivity (20 $W.cm^{-1}.K^{-1}$), high carrier mobility (>3500 $cm^2.V^{-1}.s^{-1}$ for holes and electrons) and its remarkable high breakdown electric field (>10 MV/cm) [2]. The figure of merit of diamond ranks it as one of the most promising among the conventional and wide bandgap semiconductors (Si, SiC and GaN) [3]. Several research groups are currently working on homoepitaxial diamond Schottky barrier diodes (SBDs) based on vertical or lateral-type structures [4-10]. Vertical-type structure offers a small series resistance allowing low power loss and high current density operation in the forward bias regime. However, this structure can drastically impact the reverse current, in particular when for example crystal defects at the substrate surface propagate into the active diamond layer. In this sense, Tatsumi et al. [4] correlated the increase of the reverse leakage current with the density of etch pits revealed by microwave plasma-assisted etching. In another approach based on non-destructive optical characterization, Ohmagari et al. [11] applied cathodoluminescence (CL) mapping and found a correlation between the large reverse-currents measured in certain Schottky devices and a localized band-A luminescence evidenced in the same devices. Both studies point out the crystal dislocations as the most probable electrical path through which the current can flow favoring the leakage current in the Schottky devices.

The following study builds on the work of Ohmagari et al. [11], particularly in terms of non-destructive local characterization techniques that we combined in order to deepen the understanding of the leakage current origin in vertical-type diamond Schottky devices.

## 2. Experimental details

In this study we have specifically performed mapping experiments based on conducting probe atomic force microscopy (CP-AFM) and confocal micro-Raman/photoluminescence (PL) imaging, the main idea being to electrically evidence the leakage current by CP-AFM, and perform subsequently in the same local area a structural and an optoelectronic analysis using Raman and PL. For this purpose, a confocal microscope (alpha300R from Witec) coupling in the same set-up Raman, PL and CP-AFM was used. The Raman and PL measurements were performed with a frequency-doubled NdYag laser operating at 532 nm as excitation source with up to 20 mW continuous wave output power. The spectra were recorded through a 300 mm imaging spectrometer equipped with both a 600 lines/mm and 1800 lines/mm grating, and a back-illuminated CCD. A 100x dry and oil immersion objective with respectively a numerical aperture (NA) of 0.9 and 1.4, and a 50 µm core diameter multimode fiber acting as a pinhole were used to collect the Raman and PL signal. The expected lateral spatial resolution in the x-y plane in the particular case of Raman measurements can reach 150 nm (NA=1.4) and 250 nm (NA=0.9). A particular attention was paid to the incident laser power density to avoid local heating effects that can induce a shift of the Raman peaks. No notable effect was observed even at full power, nevertheless, despite the excellent properties of diamond as heat dissipator the mapping measurements were performed with an average power of 3 mW. Local electrical measurements through the CP-AFM technique were conducted with the home-made extension called "Resiscope" [12] that was adapted to the existing AFM. This extension allows us to apply a stable DC bias voltage (from -10 to +10 V) to the sample and to measure the resulting current flowing through the tip as the sample surface is scanned in contact mode, yielding a local resistance map covering resistance values in the range $10^2$-$10^{12}$ Ω. Current-Voltage (I-V) measurements are also permitted with this extension. Highly-boron doped (HBD) diamond- and platinum silicide-coated Si AFM cantilevers, with an intermediate spring constant of about 3 N/m, proved to be the most suitable AFM tips for making electrical measurements on diamond layers. In this study the diamond active layer involved in the fabrication of vertical-type SBDs was composed of a lightly boron-doped p-type diamond homoepitaxial film grown on a HBD high-temperature high-pressure (HPHT) type-Ib single-crystalline diamond (100) substrate. The thickness of the homoepitaxial diamond layer and its acceptor concentration was estimated around 0.5 µm and $10^{15}$ cm$^{-3}$, respectively. The exact details of the growth conditions, the fabrication procedure of 150 µm diameter SBDs and the typical I-V behaviors were reported in Ref. [11].

## 3. Results and discussion

An illustration of the diamond surface with several circular SBDs is displayed in figure 1. It is important to emphasize that depending on the location of the Schottky electrodes the I-V characteristics can differ tremendously with rectification ratios in the range 1 to $10^8$ at ± 3 Volts. In what follows, we will focus only on the location areas where the SBD devices showed a high reverse leakage current (rectification ratio ~1). As an example, figure 2 illustrates a CP-AFM scan (30×30 µm²) pointing out the surface topography (left map) and the corresponding electrical image (right map) of an area composed of crystal defects formed during the epitaxial diamond growth. The topography reveals various surface defects composed of growth striations, macro-steps (MSs) and etch patterns (EPs). The associated electrical image performed with an applied voltage of 7 Volts, shows predominantly high local resistance values (pink color) except in a confined region where the local resistance decreases by six orders of magnitude (yellow color).

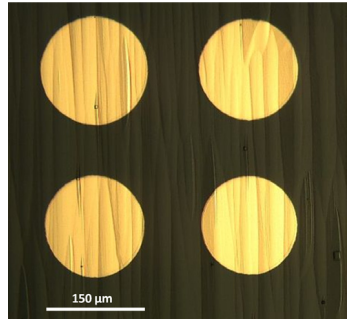

**Figure 1**. Optical microscope view illustrating part of the sample surface and the circular SBDs.

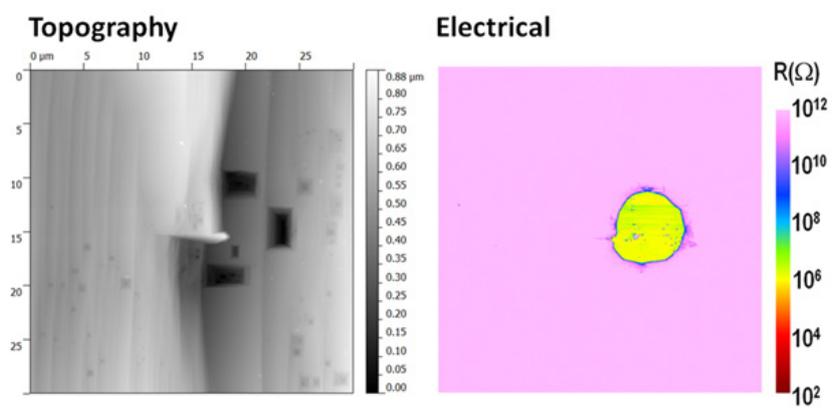

**Figure 2.** Surface topography image (left) and associated electrical image (right) of local crystal defects at the diamond surface

In order to better correlate the surface topography features and the electrical ones, we decided to stack the CP-AFM image pair into one multilayer image. For this purpose the levels of opacity of both images were adjusted to conjointly discern the topography and the electrical features. Figure 3 illustrates four examples of stacked images, including the previous figure (figure 2), measured in four different locations at the sample surface, named b7, g2, c2 and g1. The same voltage (7 Volts) was applied for those different scans. The four images reveal very similar features in terms of surface topography and local resistance. Indeed, the surface morphologies reveal growth striations (also called growth bands) and MSs whose origins are commonly attributed to the differences in lateral growth-rate and the effect of step bunching [13,14]. The latter is reported as a consequence of the misorientation of the diamond substrate and the adsorption of impurities on the diamond surfaces during the growth [14, 15]. The EPs consist of square etch pits and rectangular flat-bottomed pits whose depths can reach up to 500 nm. These etch defective areas are commonly associated to basic crystal defects, such as dislocations, impurity/cluster point defects and growth sector boundaries [2, 14]. As regards to the electrical mapping a common singularity based on a strong decrease of the local resistance (5~6 orders of magnitude with respect to the surrounding) is evidenced in a confined area delimited by the EPs and the MS corner. The following confined area will hereinafter be referred as electrical "hot-spot".

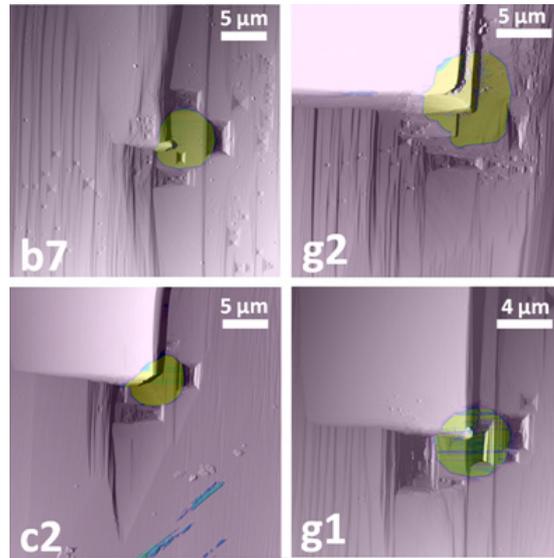

**Figure 3.** Stack of surface topography and electrical maps measured by CP-AFM in four different locations (b7, g2, c2 and g1) on the sample surface. Same voltage (7 Volts) was applied for those different maps.

Raman imaging measurements have been performed in the same locations as the previous ones illustrated in figure 3. Focusing more specifically on the location named b7, figure 4 displays four maps extracted from the spectral features of the first-order diamond Raman line at 1332 cm$^{-1}$ and the background signal. In particular, the maps show the integrated intensity, the center of mass shift (CMS) and the full width at half maximum (FWHM) of the diamond line, and the PL/fluorescence background signal, respectively.

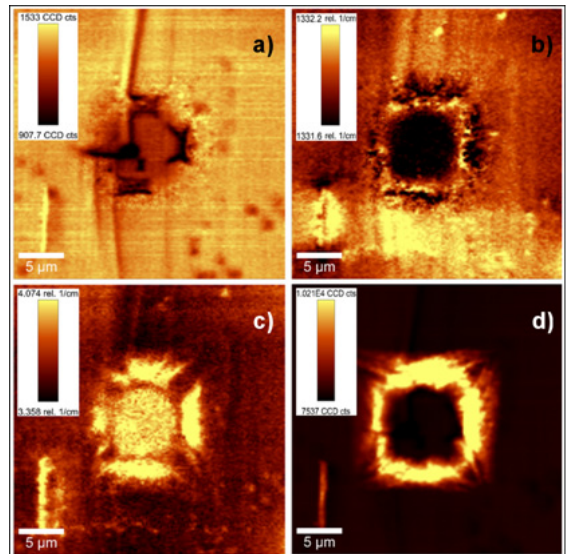

**Figure 4.** (a) Integrated intensity, (b) center of mass shift and (c) full width at half maximum of the diamond Raman line, and d) PL/fluorescence background signal measured at the location named b7.

Due to the confocality of the system and the numerical aperture of the microscope objective, the integrated intensity map (figure 4(a)) reflects quite well the topography of the surface pointing out the local crystal defects observed previously by CP-AFM (figure 2 (topography image) and figure 3

(location b7)). Especially, the EPs and the MS corner are distinguishable and delimitate an area with a weaker integrated intensity compared to the rest of the map. The same area is also highlighted by the CMS (figure 4(b)) and the FWHM (figure 4(c)) maps where a slight shift to lower wavenumbers and a slight FWHM widening of the diamond Raman line are evidenced. The PL background map (figure 4(d)) shows a noticeable signal forming a corona-like pattern, which involves the EPs and the MS crystal defect. The resulting Raman maps obtained for the location b7 can be transposed to the locations g2, c2 and g1 (not illustrated). Indeed, all the CMS and the FWHM maps reveal a singular area of a few tens of square microns (~20 µm²) characterized by a tensile stress and a low crystal quality. In addition, the corona-like pattern is also observable underlying an impurity-related PL as the most probable origin.

Figure 5 shows the Raman spectra collected from three different regions. The map inset, which is a replica of figure 4(d), illustrates these three regions labeled A (surface between the white circle and the black ellipse), B (surface inscribed in the white circle), and C (outer surface of the black ellipse). For each region an averaged Raman spectrum was obtained integrating all spectra (pixels) enclosed in the delimited region. The resulting Raman spectra were then normalized with respect to the intensity of the first-order diamond Raman line and magnified in the zone of interest. The spectrum labelled A is characterized by a broad and strong PL background signal in the range 2.3-1.9 eV. The major PL bands are a yellow band peaked around 2.15 eV observable with the naked eye and a less intense component noticeable around 2 eV. The nature of these bands may be correlated with nitrogen impurities and platelet defects which are well known present in the diamond substrate. In the same spectrum the Raman band at 1550 cm$^{-1}$ (a) could be assigned to a non-diamond sp² carbon phase [16].

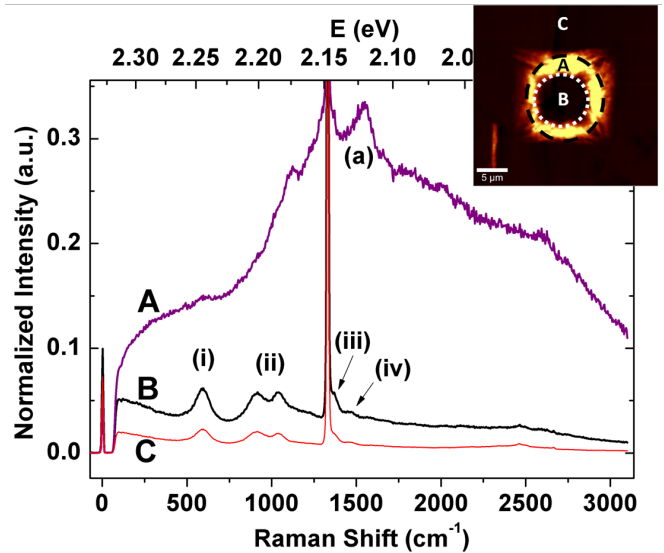

**Figure 5.** Averaged and normalized Raman spectra obtained by integrating all spectra enclosed in three different regions marked in the left map inset. The regions labeled A, B and C refer to the surface enclosed between the white circle (dotted lines) and the black ellipse (dashed lines), the surface enclosed in the white circle, and the rest of the surface map, respectively.

The spectra B and C display very similar spectral features in particular weak Raman signal bands centered at around 590 cm$^{-1}$ (i), 914 and 1040 cm$^{-1}$ (ii), their widths ranging from 80 to 150 cm$^{-1}$. In contrast to these bands, the diamond Raman line is much more intense (around 30-70 times larger) with a narrower line width (FWHM~3.4-4 cm$^{-1}$) and shows a slight asymmetry in the line shape pointing out a Fano-effect [17]. This effect is well-known in heavily doped p-type Ge and Si [18,19], but also in p-type diamond [20,21], and results from an overlapping between the inter-valence transitions and the one-phonon Raman line. The diamond Raman line also reveals a slight shoulder on its high wawenumber side at 1370 cm$^{-1}$ (iii) as well as a very weak band at 1470 cm$^{-1}$ (iv), the latter feature being ascribed to an amorphous sp2 carbon phase [16].

Comparing our spectra (B and C) to those published in several studies linked to boron-doped diamond it should be underlined that the Fano line shape and intensity observed in diamond, as well as the broad bands peaked at 590, 914 and 1040 cm$^{-1}$ may differ and change depending on the laser excitation wavelength, the laser polarization and the doping level [22-29]. In those studies a particular attention had been paid to the bands observed in the region below 1200 cm$^{-1}$ which are commonly assigned to boron impurities. Indeed, in the case of superconducting boron-doped diamond, Bourgeois at al. [23] concluded through supercell calculations and Raman experiments that boron dimers can favorably induce vibrational modes in the range 400-600 cm$^{-1}$. Additionally, it also appears that boron dimers induce empty states in the band gap which remain electrically inactive. In another study, Blank et al. [25] through polarized Raman experiments pointed out Raman bands at 588, 920 and 1042 cm$^{-1}$ for boron doping levels in the range $10^{17}$-$10^{19}$ cm$^{-3}$, they assigned these bands to phonon scattering induced by defects with the exception of the band at 920 cm$^{-1}$ attributed to the vibration of boron atoms. The same research group, Popova et al. [27], also performed calculations in case of highly boron-doped diamond (> $10^{21}$ cm$^{-3}$) and reached the conclusion that the observed bands at 580 and 860-900 cm$^{-1}$ are linked to boron localized vibrations. Based on these observations, the inhomogeneous boron incorporation could explain the band intensity variations observed between the spectra B and C in the low-wavenumber region.

Focusing to the low-wavenumber bands (590, 914 and 1040 cm$^{-1}$) and more specifically to the intensity variations, the Raman maps were reconstructed at exactly the same locations as illustrated in figure 3. Figure 6 shows the integrated intensity maps of the Raman band at 590 cm$^{-1}$ performed at the locations b7, g2, c2 and g1. The four reconstructed maps highlight a confined area that exhibits a strong integrated signal with respect to its surrounding. The integrated intensity ratio between the confined area and its surrounding can reach up to 800. Similar findings are obtained with the two other Raman bands (914 and 1040 cm$^{-1}$) after applying the same map reconstruction. It is particularly important to emphasize the analogies existing between the maps of figure 6 and those illustrated previously in figure 3, which reveal a close correlation between the highest intensity areas in terms of Raman band at 590 cm$^{-1}$ and the giant drop of the local resistance evidenced in the hot-spots.

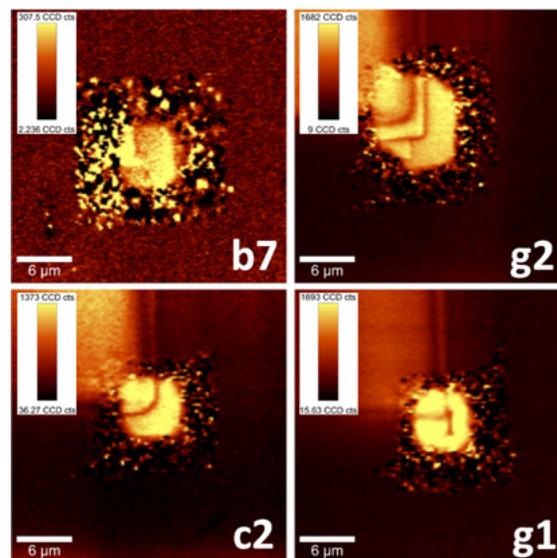

**Figure 6**. Integrated intensity maps of the Raman band at 590 cm-1 performed at the locations b7, g2, c2 and g1.

So far, all the illustrated Raman maps were collected near the surface at the same focal plane through a confocal pinhole. Additional depth information can be expected to be obtained through optical sectioning experiments, and more specifically confocal Raman depth profiling (CRDP).

Since the confocal microscope is sensitive to the depth of focus it becomes possible changing the distance sample-objective to record images with depth resolution.

The CRDP technique was performed at the same locations illustrated in figure 6, in particular x-z scans were performed in the areas involving the highest integrated Raman signal levels that were identified as the hot-spots emplacement. Figure 7 displays a Raman depth scan performed at the location named b7 with an oil immersion objective allowing an axial resolution of 380 nm. The reduced images at the top are the x-y confocal Raman maps previously shown (figure 4(a) and figure 6 (location b7)) on which a white dashed arrow has been added to localize the x-z depth scan. Figures 7(a) and 7(b) show the reconstructed depth maps obtained, as the corresponding reduced maps above, integrating the intensity of the diamond Raman line and the low-wavenumber Raman band at 590 cm$^{-1}$, respectively. The first common observation concerns the decrease of the Raman intensity as function of depth which can be understood as an effect of the absorption of the scattered light. If we compare both maps, from top to bottom, one can remark that the diamond Raman signal is detected first compared to the Raman signal at 590 cm$^{-1}$ except in the central part of the map where both signals are simultaneously observed. To guide the eye, white dotted lines have been drawn to illustrate this point. The same white lines reasonably display the position of the sample surface. The depth shift existing in certain regions between the diamond signal and the Raman signal at 590 cm$^{-1}$ serves to indirectly identify the homoepitaxial layer. Indeed, in principle the Raman signal at 590 cm$^{-1}$ is expected to be very weak in the homoepitaxial layer and, on the other hand, very intense in the diamond substrate due that is HBD. Black dotted lines have been added to the map to locate the interface between the homoepitaxial layer and the substrate.

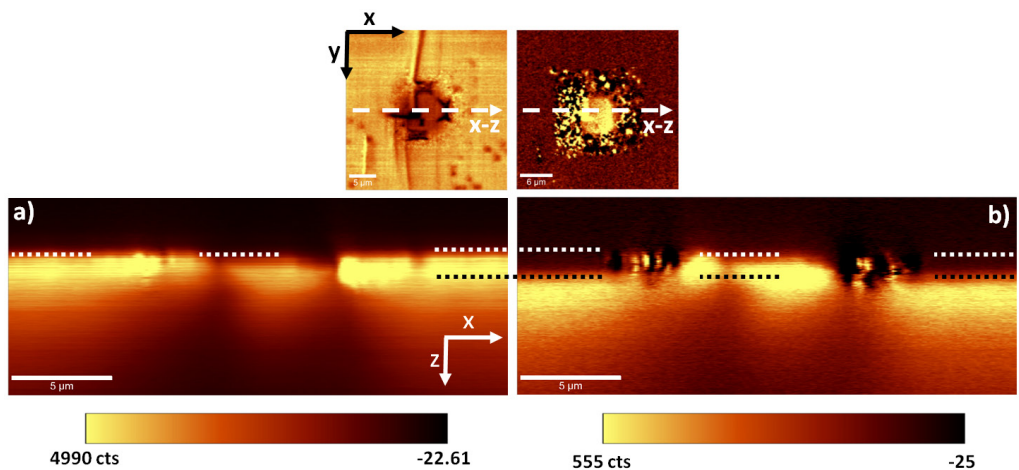

**Figure 7.** Reconstructed depth maps on the x-z plane illustrating the integrated intensity of (a) the diamond Raman line and (b) the low-wavenumber Raman band at 590 cm$^{-1}$. The reduced maps above represent the corresponding x-y Raman maps where the white dashed arrow locates the x-z depth scan.

The purpose of figure 7 is essentially to bring a qualitative description of the depth distribution of the Raman signal at 590 cm$^{-1}$ that extends from the surface at the hot-spot location to the substrate. Quantitative depth analysis is here not an easy task due that the Raman spectra collected at a given depth always contain features involving materials present above and below the focal point, and in addition the large refractive index of diamond (n=2.5 for diamond) and the point spread function (PSF) of the optical system are two critical issues that must be also take into account for a precise depth analysis.

In order to go a bit further from the electrical point of view, current-voltage (I-V) measurements were carried out locally with the CP-AFM set-up. More specifically the measurements were performed in the hot-spots identified in the locations b7, g2, c2 and g1. Due to the symmetry of the I-V curves only the characteristics for a positive applied voltage are plotted. Figure 8 shows four series of I-V curves measured with a HBDD AFM tip, the curves are illustrated on a log-log scale for clarity. Indeed, for V > 0.2 V all the curves clearly display a linear behavior pointing out a power law relationship, $I \propto V^m$, where m refers to the slope which reaches values close to 3. Conversely, the I-V trends below 0.2 V are not clearly identified. Nevertheless, trap-controlled space-charge-limited current (SCLC) remains the determinant transport mechanism that justifies the trap-filled-limited (TFL) conduction regime observed for V > 0.2 V. TFL conduction mechanism can be understood as the result of the carrier injection that saturates all trap defects. Indeed, after the trap saturation the carrier injection accumulates in the allowed band provoking a steep increase of the current with a slope higher than 2. SCLC mechanisms in diamond have been described by several authors and used as an electrical method for probing the density of states above the valence band [30-34]. Some of these studies concluded with the presence of an exponential distribution of traps that extends up to the valence band.

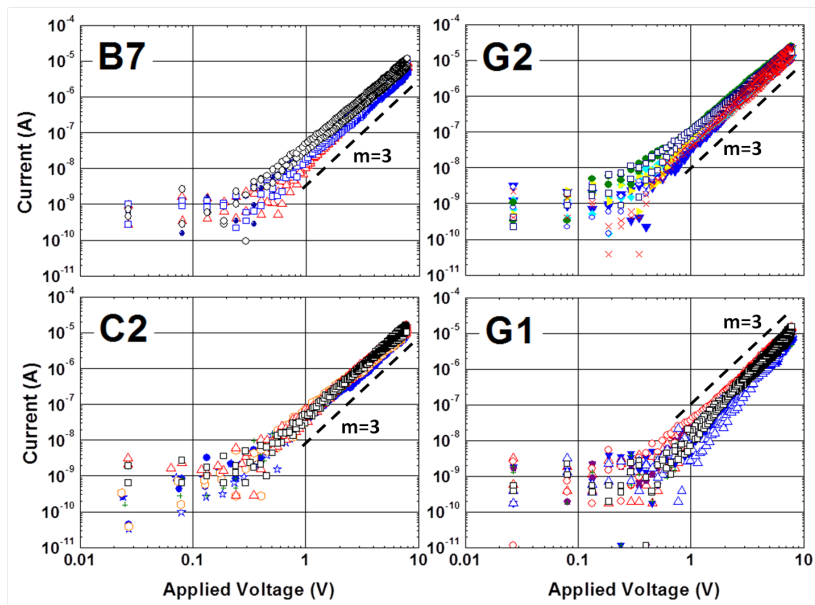

**Figure 8.** I-V curves performed by CP-AFM in the hot-spot areas at the locations b7, g2, c2 and g1. Dashed lines showing a slope of 3 are drawn to guide to the eye.

Based on the following results, the first remark that can be made is that the high leakage current observed in several Schottky diodes is controlled by TFL conduction mechanisms taking place at the hot-spots. However, this implies that the metal used as a Schottky barrier behaves locally as an injecting point contact when in particular it interacts with the area of the hot-spot. The identification of the trap defects (point defects, dislocations, impurities and their possible complexes...) involved in the TFL conduction mechanism remains a difficult task, however in view of the above Raman results the boron impurities and its possible complexes might be strongly put forward. Boron atoms are in diamond the most successful p-type dopant whose concentration allows transitions from semiconducting to metallic and to superconducting properties. Nevertheless as investigated by Muret et al. [35] and Elsherif et al. [36], some of the incorporated boron atoms do not act as acceptors, but rather as deep level traps. Some of these levels based on boron complexes are proposed to be due to the segregation of boron atoms into the defective areas (point, linear and planar defects) that were created or propagated during the diamond growth. In this regard, the Raman maps of figure 4 ((a), (b) and (c)) highlight a local area where the diamond lattice appears to be more constrained. In addition, the same local areas are revealed by integrating the intensity of the low-wavenumber Raman bands (590, 914 and 1040 cm$^{-1}$) as displayed in figure 6. The link between these Raman bands and the electrical hot-spots are clearly established when one compares the reconstructed Raman maps in figure 6 and the CP-AFM maps in figure 3. Indeed, the area of the hot-spot coincides with an increase of the integrated intensity of the Raman bands at low-wavenumber which are commonly attributed to the incorporation of high amount of boron into the diamond lattice. Lastly, the depth Raman maps of figure 7 point out in a qualitative manner that the hot-spot is located in the homoepitaxial layer, and that the low-wavenumber Raman bands characterizing the hot-spot are also clearly identified in the HBD diamond substrate.

The close correlation found by Ohmagari et al. [11] between the band-A luminescence (assigned to the presence of dislocations) and the high leakage current is here not observable due to our experimental conditions which do not allow the excitation of the blue band-A (~ 2.8 eV). However as illustrated in many studies [37-39], the incorporation of boron in diamond layers also induces a broad green luminescence (~ 2.3 eV) which overlaps the band-A emission and can become the prominent PL band in case of high boron concentration. The intensities of both bands are also dependent on the growth sectors due that the defects/impurities propagate/incorporate themselves in a different way during the growth phase. The following remarks are in some way consistent with the results presented in this study given that the area of the hot-spots is constrained, confined in a region delimited by crystal defects (MS and etch pits) and boron-rich.

## 4. Conclusion

In conclusion, the origin of the high leakage current in vertical-type diamond Schottky devices was investigated through non-destructive local characterization techniques based on CP-AFM and confocal micro-Raman/PL imaging analysis that were performed in the same local areas. Electrical hot-spots of a few tens of square microns in size have been identified and correlated, in a first stage, with the constraints in the diamond lattice. However, a thorough analysis of the integrated intensity of the low-wavenumber Raman bands clearly established a close link between the electrical hot-spots and the high-amount of boron locally incorporated. The major unknown concerns how the boron is incorporated into the lattice and how it acts electrically. In this sense, I-V characteristics performed at the hot-spots showed a trap controlled space-charge-limited current as the main conduction mechanism that is responsible for the high leakage current observed in the SBD devices.